\documentclass[letterpaper, 10 pt, conference]{ieeeconf}  

\IEEEoverridecommandlockouts                              
\usepackage[utf8]{inputenc}
\usepackage[T1]{fontenc}
\usepackage{graphicx}
\usepackage{amsmath}

\title{\LARGE \bf
Simplified markerless stride detection pipeline (sMaSDP) for surface EMG segmentation
}

\author{Rafael Castro Aguiar$^{1}$, Edward Jero$^{2}$ and Samit Chakrabarty$^{1}$
\thanks{$^{1}$R. Castro Aguiar and S. Chakrabarty are with the School of Biomedical Sciences, Faculty of Biological Sciences,
        University of Leeds, Leeds LS2 9JT, UK,
        {\tt\small email: bsrdca@leeds.ac.uk}}%
\thanks{$^{2}$S. Edward Jero is with the Indian Institute of Technology Madras. Chennai 600036, India, 
        {\tt\small email: edwardjero@gmail.com}}%
}

\begin{document}

\maketitle
\thispagestyle{empty}
\pagestyle{empty}

\begin{abstract}

People with mobility impairments are often recommended for gait assessment studies to diagnose their condition and to select appropriate physiotherapy to improve their mobility. These studies are often conducted in clinical or lab settings, where subjects are assessed in a foreign environment, which may influence their motivation, coordination and overall mobility. Alternatively, if the subject's gait could be assessed in their daily-lives, in unconstrained settings, a more naturalistic gait assessment could be performed. Kinematic analysis of a gait pattern on its own may not be sufficient to characterise a subject's mobility. To better diagnose gait deficiencies, analysis of the patient's muscle activity should be conducted as well. To do so, gait studies should collect, synchronously, Electromyography (EMG) and kinematic data.

This method introduces a simplified markerless gait event detection pipeline for the segmentation of EMG signals, via synchronously recorded Inertial Measurement Unit (IMU) data. In an unconstrained walking experiment, healthy subjects walk through a designed course with their kinematic and EMG data recorded. This course comprises 5 different walking modalities (level walking, ramp up/down, staircase up/down), mimicking everyday walking. Through timepoint matching, segmentation and filtering, we generate an algorithm that detects heel-strike (HS) events using a single IMU, and isolates EMG activity of gait cycles, in the different walking modalities.

This gait event detection algorithm can be adapted to different datasets, and was tested in both healthy and Parkinson's Disease (PD) gait. Results demonstrate the extracted muscle activity levels in a healthy subject's level ground walking, and the extracted HS events of a PD patient. Adjustments to algorithm parameters are possible (e.g., expected velocity, cadence) and can further increase the detection accuracy.

\end{abstract}
\section{INTRODUCTION}

Investigation of gait patterns is a valuable diagnostic tool that provides clinicians the ability to diagnose a range of impairments affecting a patient's mobility, such as trauma, stroke or neurological conditions. Features extracted from these gait patterns, as key-event times, duration, and joint-angle deviations are all quantifiable measures to allow a clinician to benchmark and make informed decisions over the selection of appropriate therapies.

\begin{figure}[h!]
      \centering
       \includegraphics[width=0.45\textwidth]{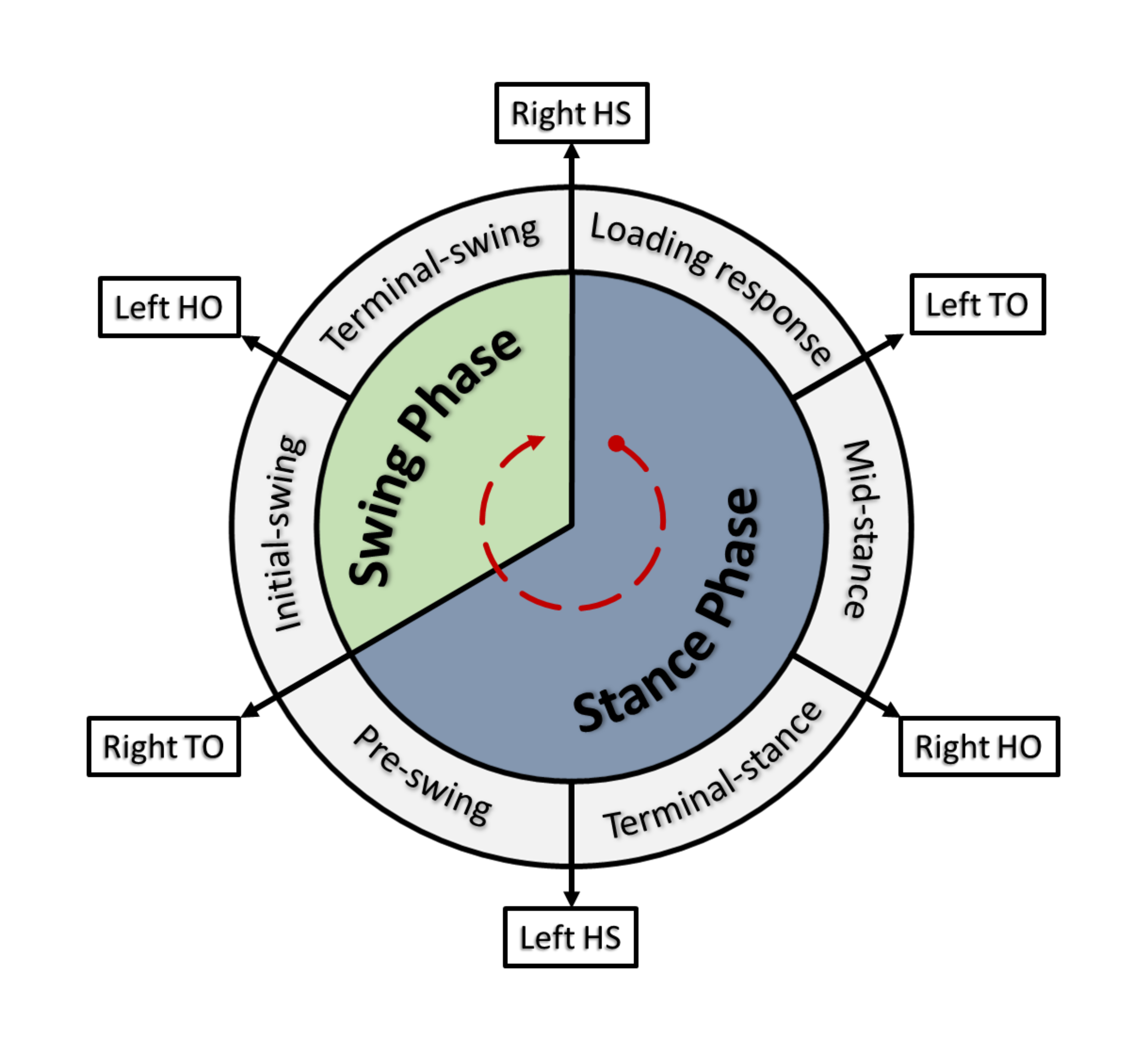}
      \caption{Simplified stages of a typical gait cycle, adapted from \cite{whittle2007gait}, for a right leg dominant subject. A new cycle is initiated with the \textbf{right heel strike (HS}) and with both feet in contact with the floor. Following a loading response, the left foot is lifted off the floor after the \textbf{left Toe-off (TO)}, and weight begins to shift on the right foot from heel to mid-foot as the body moves forward. After mid-stance, the right heel loses contact with the floor in the \textbf{right Heel-off (HO)} stage, and weight shifts towards the right forefoot. In preparation for the right limb swing phase, the left foot once more contacts the floor with a \textbf{left heel strike (HS)}. With double support on the floor, the right foot enters the swing phase, following the \textbf{right Toe-off (TO)}. The same stance process now happens on the left foot, with weight shifting from heel to toe, and respective \textbf{left Heel-off (HO)}. This current gait cycle finally concludes with the right HS, and a new cycle initiates.}   
      \label{GC_stage}
\end{figure}

In essence, a gait cycle is comprised of two core phases for each limb: stance and swing. This phase differentiation is associated with foot contact with the floor (stance) and swinging motion of the limb through the air, preceding the following step (swing). Taking the example of a right leg dominant subject, Figure \ref{GC_stage} describes the gait cycle for a naturalistic walking pattern. For more detailed explanations of the gait cycle, please refer to \cite{whittle2007gait,tao_gait_2012,gujarathi_gait_2019}.

Typical gait analysis relies on a trained clinician's visual and qualitative assessment. To aid in the visual inspection, gait analysis often makes use of high-cost, complex optical recording setups with associated force plates, installed in a clinic or laboratory space. These systems record, through a combination of high-speed cameras, the individual's movement via tracking of a predetermined set of retroreflective markers placed on the individual's body. Although accurate and reliable, optical motion capture systems restrict gait studies to confined locations within the camera setup, and incur in high training requirements for examiners to setup, calibrate, collect and analyse gait data. Due to this lack of versatility, optical systems are unable to investigate a patient's daily life mobility. To tackle this issue, researchers pushed towards the development of wearable technologies, through sensors that could still reliably identify gait events, without the typical constraints of optical systems. 

To best identify which are the most appropriate sensors for gait event detection on Activities of Daily Living (ADLs), and to pair with EMG analysis, a short review and selection of available sensors technologies was conducted. One of the earliest reported wearable technologies for gait event detection uses foot-switches \cite{WINTER1972553}. These sensors are usually embedded into the shoe or insole, located underneath the heel and hallux of the individual. This allows for the recording of two key gait events: Initial-Contact (IC), in healthy scenarios reported as the heel strike (HS); and End-of-Contact (EC), or Terminal Contact (TC), in healthy scenarios reported as the toe-off (TO). With current iterations, as insole pressure sensors (IPS) and force sensitive resistors (FSR), fitted into footwear (e.g. \cite{hanlon_real-time_2009,han_gait_2019,nazmi_walking_2019}), this technology is widely employed today and used as a benchmark for other motion capture systems (see \cite{prasanth_wearable_2021} for an up-to-date review). A common issue with these IPS and FSR systems is the constant wear-and-tear of the sensors due to associated shear-forces in the sole, or against the walking surface. This leads to short sensor lifespans and to sensing accuracy loss after sustained periods of usage.

Nowadays, the most commonly employed wearable technology for gait event detection is accelerometry (\cite{prasanth_wearable_2021}). First described in \cite{MORRIS1973729}, this method makes use of acceleration sensors, strategically placed near subjects joints on predetermined sites. The choice of placement will greatly influence the gait parameters extracted and takes into consideration possible noise motion artefacts that contaminate the recordings (e.g. poor sensor adhesion or loose clothing).
With advances in wearable technology, typical accelerometer sensors were adapted into IMUs (Inertial Measurement Units), sensors that are able to detect both acceleration and angular velocity (through an integrated gyroscope) in all three dimensions. Depending on the desired gait parameters, the IMU sensors are most often placed at the foot (\cite{jasiewicz_gait_2006,hanlon_real-time_2009,tadano_three_2013,nwanna_validation_2014,ju_pedestrian_2016,park_stance_2016,zhao_imu-based_2017,mei_determination_2019,zhen_walking_2019,perez-ibarra_real-time_2020,liu_novel_2021,wouda_foot_2021}) or shank/calf (\cite{jasiewicz_gait_2006,tadano_three_2013,grimmer_stance_2019,gujarathi_gait_2019,han_gait_2019,zhen_walking_2019,romijnders_validation_2021}), but other studies present interesting alternatives as knee/thigh (\cite{torrealba_characterisation_2007,hanlon_real-time_2009,wentink_intention_2013,ryu_semg-signal_2019,grimmer_stance_2019}, waist (\cite{yang_real-time_2011}) or even head (\cite{hwang_real-time_2016,park_validation_2021}) placements.
During the aforementioned gait events (i.e. HS and TO), there are distinguishable peaks in the vertical acceleration profiles that can be used to identify the timing of HS and TO. Upon contacting the surface, the heel will experience a vertical force (ground-reaction force) and its respective upward acceleration, which is identifiable when inspecting the recorded signal.
In addition to gait event detection, IMUs also presents benefits in the detection of angular displacements and velocities of different joints (\cite{wouda_foot_2021,han_gait_2019,zhao_imu-based_2017}) as well as pedestrian tracking solutions (\cite{ju_pedestrian_2016}). With state-of-the-art sensor synchronisation, current IMU solutions also provide accurate position estimation for the subjects wearing the sensors, as described in \cite{schepers_xsens_2018}. 
Due to all these features, comparison studies (\cite{tadano_three_2013,prasanth_wearable_2021,han_gait_2019,nazmi_walking_2019,wouda_foot_2021,hanlon_real-time_2009}) often place the IMU as a preferable wearable solution over the previously discussed alternatives (i.e. FSR and Optical systems). Even if the IMU is not as accurate as the FSR or as thorough in limb position detection through optical systems, the IMU sensor durability, robustness, cost and applicability outweigh the benefits of these other technologies in ADL gait detection applications.

So far, the wearable technologies presented here rely on external interactions of the individual with the sensors (e.g. pressure applied on the IPS) or measurements of external inferred parameters (e.g. the body part acceleration or angular velocity as measured by the IMU) to identify key gait events. An alternative is to make use of the patient's biological signals, as muscle activity recordings through surface Electromyography (sEMG), to estimate the timing of the events. sEMG provides a reliable non-invasive solution to analyse muscle activity levels during varied motor tasks, and in a non-constrictive manner.
sEMG can be used to diagnose motor activity, where the recorded muscle activity is analysed to identify muscle deficiencies, and/or establish individual muscle function levels during pre-defined tasks.
This has been specifically used for the analysis of gait, using analysis of the sEMG recorded from hip, knee and ankle flexor and extensor muscles, reported in \cite{tao_gait_2012,wentink_intention_2013,suwansawang_analysis_2018,morbidoni_deep_2019,nazmi_walking_2019,ryu_semg-signal_2019} with positive evidence for differentiating activity in different segments of the gait cycle. However, it being a biological signal, the sEMG can be quite noisy and variable, depending on sensor location and other biological factors.

To interpret muscle activity features and classify these into gait events, studies often return to the use of machine learning algorithms (\cite{wentink_intention_2013,morbidoni_deep_2019,nazmi_walking_2019,ryu_semg-signal_2019}). Unfortunately, as mentioned, there are several factors that may interfere with the sEMG quality and the accuracy of these classification algorithms, especially when recognising how volatile and different the recorded signals can be across different individuals, and even in different muscles of the same individual.

All these wearable solutions present promising results for gait event detection, through varied technologies and algorithms. The complexity of such tools decides how viable it is for applications of gait assessment during Activities of Daily Living (ADLs). Unfortunately, these solutions seldom come with a clear methodology of how to apply these tools to detect the most simple of gait events, rarely offering the algorithms to do so. In this method paper, we propose a simplified method of heel strike detection for synchronised sEMG recording segmentation through IMU data paired with sEMG recordings, applied to the diverse and unconstrained dataset of \cite{brantley_full_2018}, and further test it with samples of data from \cite{PD_gait2010}, looking at Parkinson's Disease (PD) gait.

\section{METHODOLOGY}

The following section details the steps, and associated scripts, to isolate sEMG activity patterns in different unconstrained gait modalities, based on the kinematic dataset of \cite{brantley_full_2018}. This guide contains step-by-step information about using recordings from an IMU system (\cite{xsens_mvn_2021}), focused on a single sensor on the subject's dominant leg ankle, to identify heel strike timestamps during different gait patterns. For ease of processing, each important step of the sEMG segmentation is tackled and discussed in separate scripts, coded in \textit{MATLAB}(\cite{matlab2020}). 

In summary, the dataset provided in \cite{brantley_full_2018} contains unconstrained walking data from 10 healthy participants, with kinematic (through IMU system \cite{xsens_mvn_2021}), sEMG and EEG (Electroencephalography) recordings. The participants walk through a designed walking course, encompassing a ramp, level ground and staircase sections. In each trial the participant walks up and down the ramp and staircase, as well as level ground walking, providing recordings for 5 different walking modalities. For ease of referencing, please consider the following acronyms: 

\begin{itemize}
    \item \textbf{RA}: ramp ascent;
    \item \textbf{RD}: ramp descent;
    \item \textbf{SA}: staircase ascent;
    \item \textbf{SD}: staircase descent;
    \item \textbf{LGW}: level ground walking;
\end{itemize}

\begin{figure}[h!]
      \centering
       \includegraphics[width=0.48\textwidth]{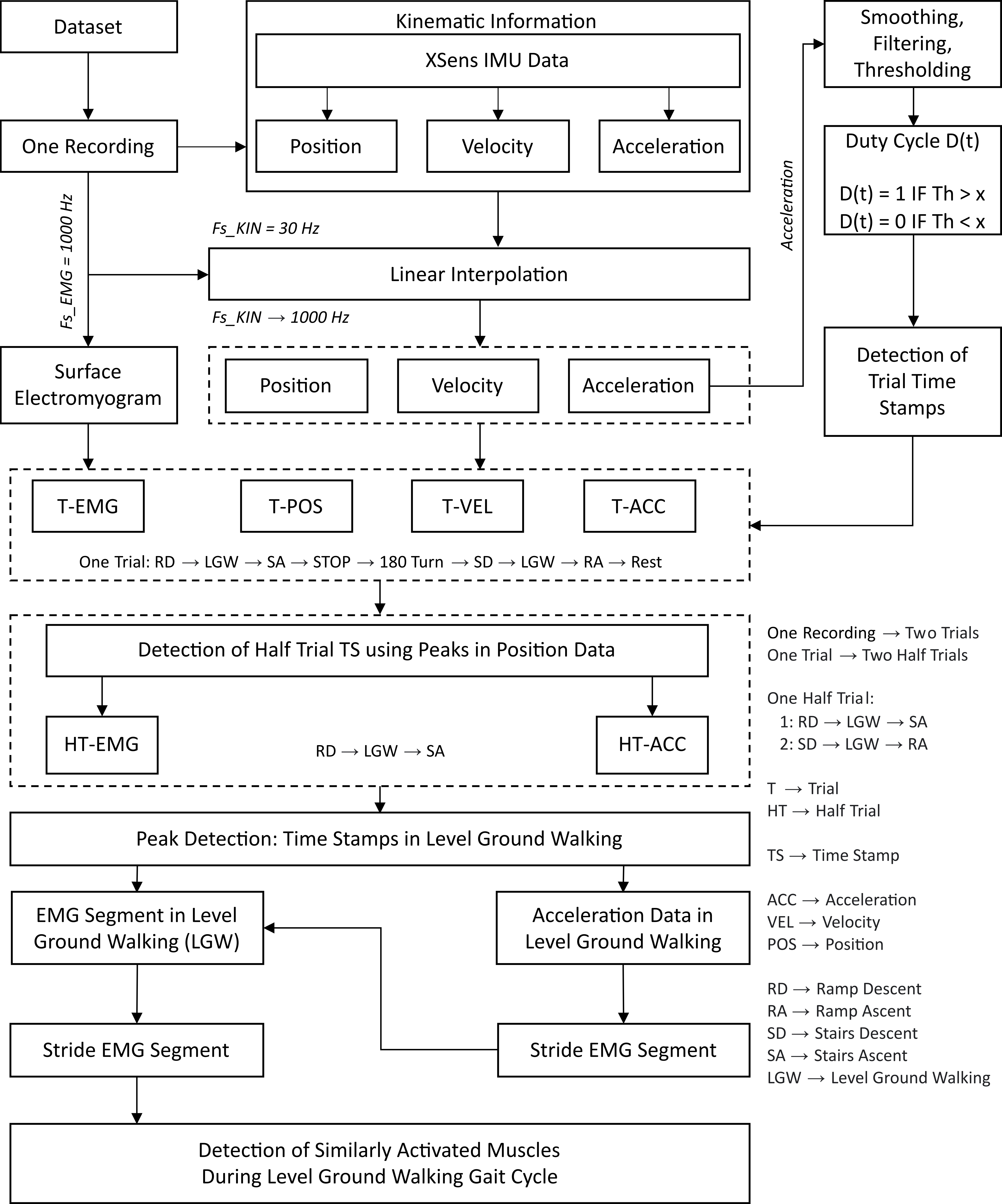}
      \caption{Summary of method architecture: the diverse multi-step process, from the recorded IMU signals, filtering of movement periods, segmentation into different walking modalities and identification of HS moments}
      \label{SArch}
   \end{figure}

To accompany and summarise the following methodology, Figure \ref{SArch} details the required steps and processes for the kinematic and sEMG data.

\subsection{Interpolation and kinematic activity segmentation}

Kinematic data (KIN) and sEMG data are recorded at different sampling rates, but synchronised in time. To use kinematics as an indicator for segmentation timestamps in the sEMG counterpart, the kinematic signal requires linear interpolation to match the number of sEMG datapoints.
\begin{eqnarray}
        {\frac {y-y_{0}}{x-x_{0}}}={\frac {y_{1}-y_{0}}{x_{1}-x_{0}}} \\
         y=y_{0}+(x-x_{0}){\frac {y_{1}-y_{0}}{x_{1}-x_{0}}} \\
         y={\frac {y_{0}(x_{1}-x)+y_{1}(x-x_{0})}{x_{1}-x_{0}}} 
         \label{eq:lin_int1}
\end{eqnarray}

\begin{equation*}  
    y_s(i) = \frac{1}{2N + 1} (y(i+N)+y(i+N-1)+ \\
    \label{eq:lin_int2}
\end{equation*}

\begin{equation}
    ...+y(i-N))\\
    \label{eq:lin_int3}
\end{equation}

Equations \ref{eq:lin_int1} and \ref{eq:lin_int3} present a common linear interpolation. With a similar time duration between recordings but different sampling frequencies, the kinematic signal (y) with less samples is interpolated to match the number of samples of sEMG signal (x), estimating missing kinematic values. This process ensures that both KIN and sEMG recordings are synchronised in time and samples. A moving average smoothing filter is then applied to KIN to remove unwanted oscillations and initial motion noise from the signal.

The pre-processed signals now consist of two complete walk trials that includes all five walking modalities previously discussed (RA, RD, SA, SD and LWG) for each subject. A minimal energy threshold and window are created to isolate and timestamp meaningful movement identified using the IMU acceleration. Any signal fragment discovered inside a passing window that is less than a certain energy level is categorised as '0'. Similarly, if significant acceleration is generated beyond the threshold regarded as evidence of the subject walking, it is recorded as 1. As a result, a binary signal is generated identifying datapoints of relevant walking activity.

\subsection{Filtering small movement artefacts and separating trials from a single recording}

The core objectives for this section are: filtering movement and/or transition artefacts, considered as noise; to identify the direction of movement, as the walking modalities presented are dependent on the direction of the walking course. The output of the previous section is a duty-cycle type wave, depicted as the grey square wave in Section A of Figure \ref{SDiag}), presenting the binary values of 1 when the subject is active, and 0 when the subject is standing still. The start and end of active periods define the timepoints at which walking starts and ends. As the participants complete the trials in a continuous manner without taking unnecessary breaks between walking modalities, any detected activity under a threshold of 6000 datapoints (or 6 seconds) is classified as a motion artefact or noise.

In the experiment, each recording is comprised of 2 trials. In a single trial a participant completes the course in a forward direction, rests at a designated location (B), and then returns to the starting point (A), walking in the reverse direction of the course. Each of the walking directions (A-B and B-A) is defined as a "half-trial". A single recording has then 4 half-trials, where 2 are the participant walking from the starting point (A) to point (B), and 2 where the participant is walking the reverse direction from point (B) to point (A).
Four moments of activity are then extracted from the complete recordings and stored as an independent variable to later use to define the direction in which the participant walked the course.

\subsection{Initial sEMG segmentation and direction, using the timepoints defined by the 4 moments of activity per complete trial}

The extracted timestamps for the start and end of each half-trial are then used to segment the sEMG and KIN signals. 
The sEMG recordings provided in this dataset are in raw format. The objective of this stage is to remove noise from movement artefacts and possible recording interference, and to isolate relevant frequency content from the data. For this purpose, a 4th order Butterworth bandpass filter is designed, to filter any content outside the 10 to 150 Hz band. Additionally, a notch filter is also used to remove powerline interference at 60 Hz.

The following objective, as mentioned, is to cut both sEMG and KIN signals, using the previously defined half-trial timestamps. For this purpose, and highly dependent on the dataset in question, a timestamp with a safety margin of 2000 datapoints (2 seconds) is used to ensure the inclusion of all relevant activity. sEMG and KIN are then segmented into bins of activity based on half-trials, for all trials and all participants.

As mentioned, per complete trial, 4 moments of walking activity are detected (example in Section A of Figure \ref{SDiag}), related to walking the course twice in the forward and twice in the reverse directions, as described previously between (A) to (B). Based on the extracted timestamps and the experiment methodology, the sequence of modalities in the forward direction is separated into RD, LGW and SA, and for the reverse course direction, separated into SD, LGW and RA.

\subsection{Walking modality identification within each direction of the course and trials}

\begin{figure}[!ht]
      \centering
       \includegraphics[width=0.48\textwidth]{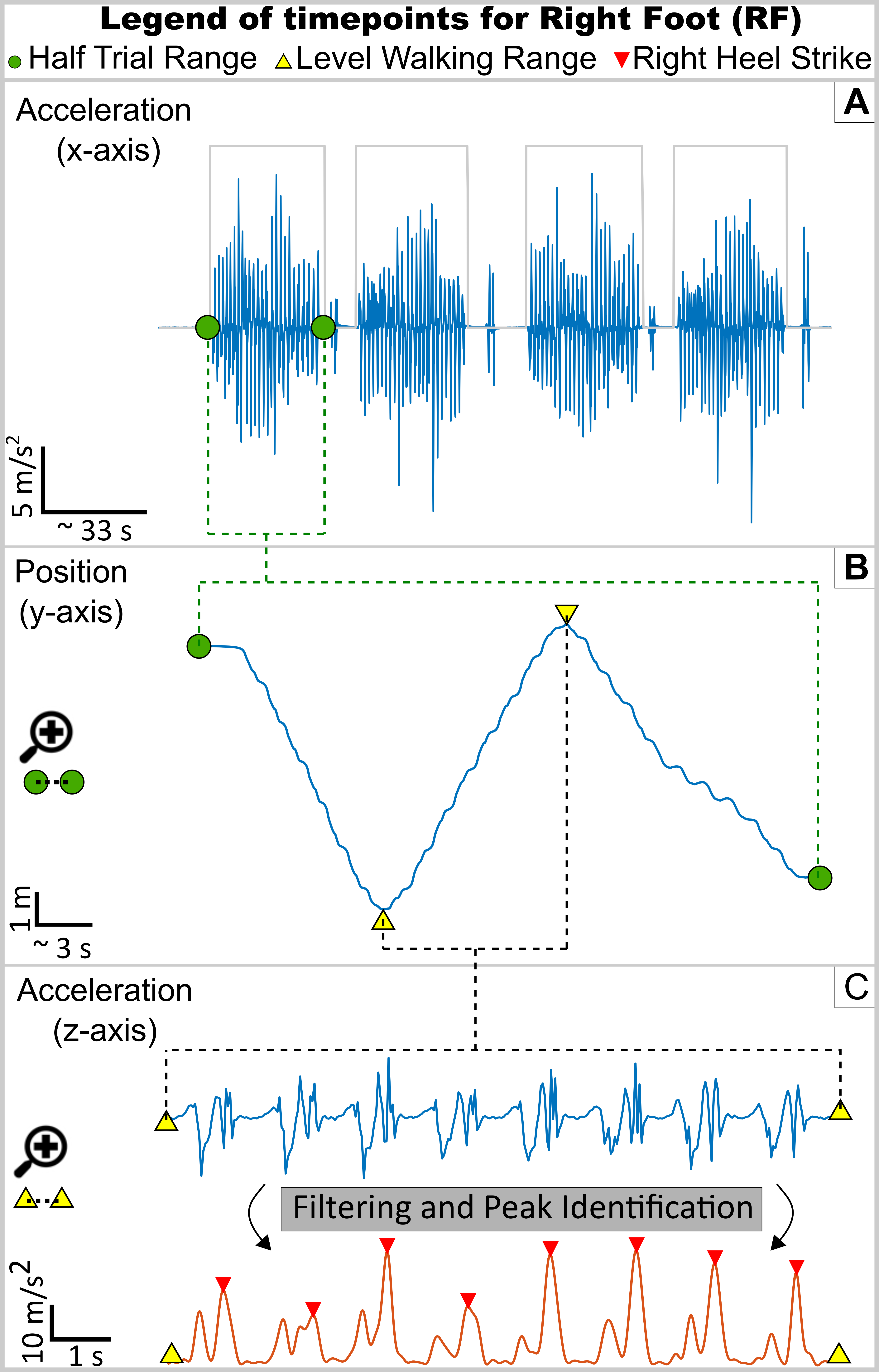}
      \caption{Summary diagram illustrating the different stages of the method, for the duration of a HT (RD,LGW,SA): in Section A, the x-axis acceleration of the right-foot IMU is used to identify the initial HT start and end timepoints, marked with the green circles; within the time range defined by the HT green circles, Section B now takes the y-axis position data from the IMU to identify the sharp direction turns related to switches in walking modality, represented here by the yellow triangles; lastly, within the isolated LGW modality between the yellow triangles, Section C uses the z-axis acceleration from the IMU, and through our filtering and peak identification algorithm, identifies the HS events, marked by the red triangles.}
      \label{SDiag}
   \end{figure}

At this stage, sEMG and KIN signals for all trials are segmented and separated according to the direction of the course the participant is walking along. The next step is to identify the timestamps for each walking modality within the forward and reverse courses, and isolate these in both the sEMG and KIN signals. Due to the design of the course, participants take a sharp turn at the transition between each modality (example on Section B of Figure \ref{SDiag}). This is identified via the estimated position changes as recorded by the IMU. Focusing on the y-axis, any time there is a peak transition, a new modality begins, according to the course guide mentioned earlier. This, paired with safety margins, is then used to further segment both sEMG and KIN into the individual walking modalities. 

\subsection{Using acceleration data to identify the moment of heel strike}

Having the data segmented into different walking modalities, the next step is to identify gait cycles within each walking modality. Through acceleration data in the vertical direction (z-axis), the moment of HS can be identified. To do so, new filters are designed to isolate the vertical peaks of acceleration witnessed when the heel strikes the ground, and experiences the ground-reaction force. Two separate 7th order Butterworth filters are created: a high-pass at 9 Hz and a low-pass at 6 Hz. These remove a band where specific motion artefacts lie, and isolate the striking moments. Being post-hoc, these filters are ideal and do not impose any delays on the data. 

The filtering process is as follows:
- the high-pass filter is applied to the raw KIN data, removing frequencies below the 20 Hz band, and potential movement artefacts;
- a half-wave rectifier is applied to the resulting filtered data, removing unnecessary negative electrical oscillations imposed by the sensor readings;
- a low-pass filter is applied to the rectified data, serving as an anti-aliasing filter at 5 Hz, to remove oscillations from the signal.
 
The HS moment is identified as the highlighted peaks in the Filtering and Peak section of Figure \ref{SDiag}. Extracting the timestamps of each of these peaks allows then for the isolation of a complete stride, or gait cycle, from one heel strike moment to the immediate next of the same limb. These timestamps are then finally used to segment the sEMG activity into separate gait cycles.

\section{RESULTS}

Figure \ref{SDiag} demonstrates the algorithm effectively identifying the moments of HS, exemplified in that segment of level ground walking. For the other walking modalities, there is a noticeable shift in the amplitude of the ground-reaction force acceleration and an adjusted cadence, but each HS is equally as recognisable using our algorithm. 

\begin{figure}[h!]
      \centering
      \includegraphics[width=0.48\textwidth]{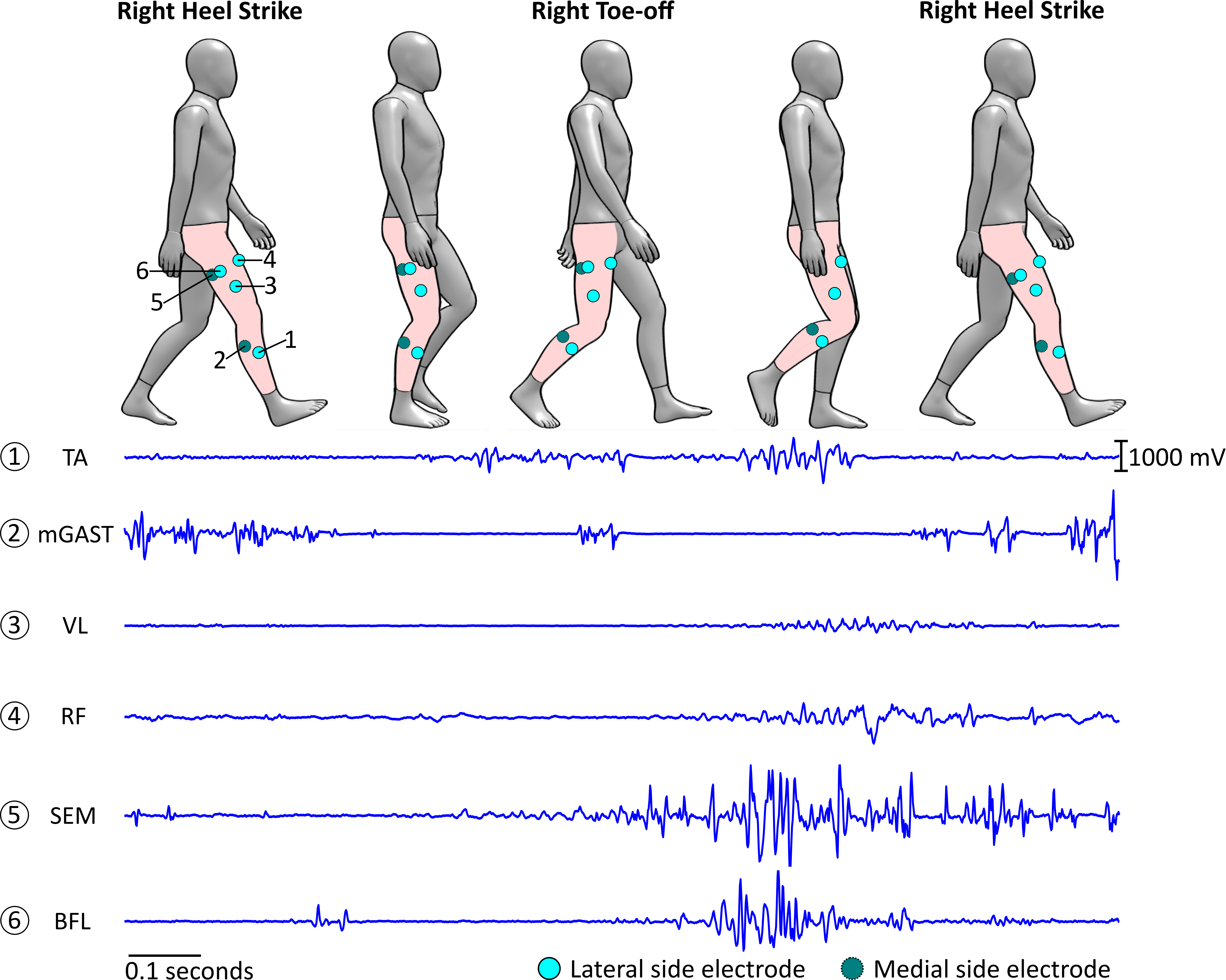}
      \caption{Segmented sEMG from a single extracted gait cycle: top section diagram represents a full gait cycle, as initially described in \ref{GC_stage}, with the respective locations of the sEMG sensors; the bottom traces are the segmented sEMG recordings for that particular cycle, with recordings from Tibialis Anterior (TA), Medial Gastrocnemius (mGAST), Vastus Lateralis (VL), Rectus Femoris (RF), Semitendinosus (SEM) and Biceps Femoris Longus (BFL)}
      \label{segEMG}
\end{figure}

\begin{figure}[ht!]
      \centering
      \includegraphics[width=0.48\textwidth]{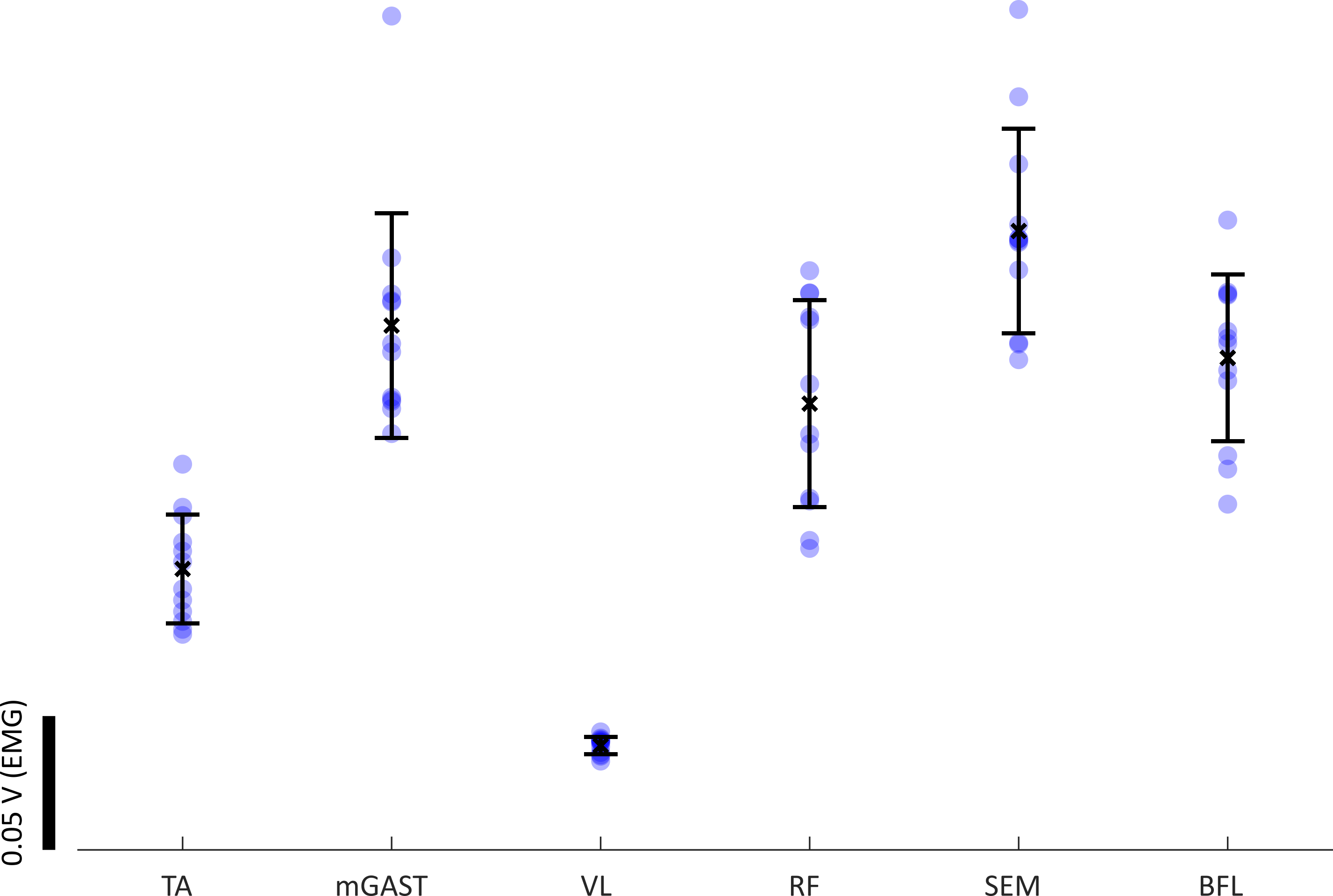}
      \caption{Observations, with mean and standard deviation of muscle activity extracted from 12 LGW gait cycles from a randomly selected trial and participant. Analysing this activity in different walking modalities allows for quantification of the variability in each muscle and the impact on the person's mobility.}
      \label{MAPs}
\end{figure}

With the final synchronised sEMG and KIN signals, individual gait cycles were then extracted and separated. In Figure \ref{segEMG}, an example of sEMG activity is presented for the same level ground walking segment of Figure \ref{SDiag}. Within the isolated gait cycles, we can identify the muscle activity patterns from the segmented sEMG data, as seen in Figure \ref{MAPs}.

As a proof of concept, this algorithm was applied to another unconstrained walking dataset \cite{PD_gait2010}, to a PD patient, performing a series of walking tasks in a therapy scenario. In this dataset, patients were using an IMU placed on their dominant foot, akin to the experiment focused on in this report. Unlike the core dataset though, this PD data does not provide sEMG recordings. Nevertheless, with fine tuning of cadence and velocity parameters, through adjustments of the peak detection section, all the moments of HS are detected over the entire trial (over 1 minute).

\begin{figure}[h!]
      \centering
       \includegraphics[width=0.48\textwidth]{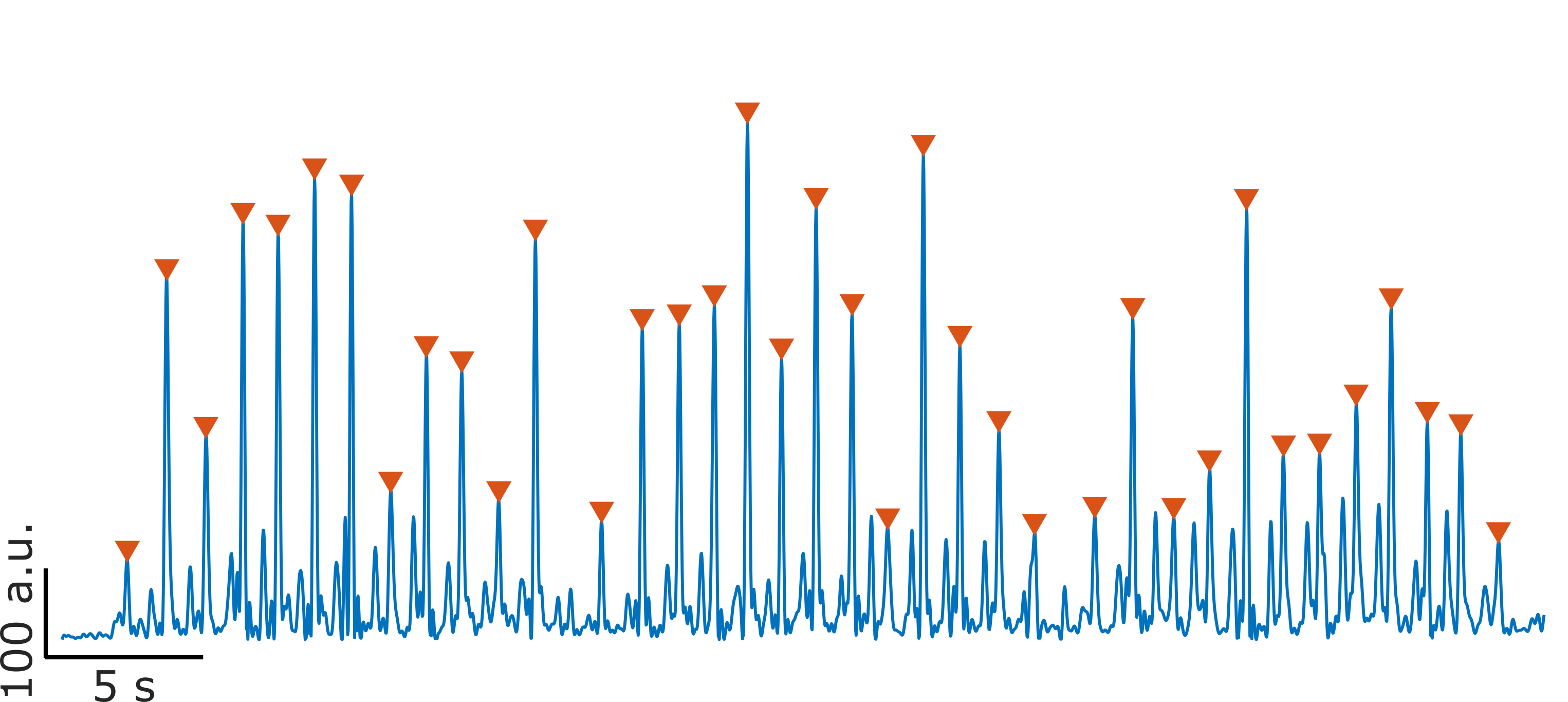}
      \caption{HS detection on unconstrained Parkinson gait, obtained from the dataset of \cite{PD_gait2010}. Recorded data represents a significant portion of time and activity of the patient, walking around a clinical room, accompanied by therapists.}
      \label{PD_gait}
\end{figure}

\section{DISCUSSION AND CONCLUSIONS}

Several methods have been designed for gait event detection, based on an assortment of different technologies, of different complexities (refer to \cite{prasanth_wearable_2021} for an up-to-date review). With advancements in sensor design it is now possible to bring these tools to our everyday lives, and gait assessment to activities of daily living. The importance of tracking patients' performance levels at their own homes and through their daily lives is of massive benefit for the design of smarter continued therapies. With the ambition of continued, supervised at-home therapies and improved rehabilitation, it is essential that researchers are provided with simple guides on using the technology and appropriate methodology to provide them the foundation for exploring therapeutic mobility.

Even in the ADL scenario, as demonstrated, kinematics alone may not provide sufficient information to therapists to fully assess patients' mobility. Integration of muscle activity levels, associated with recorded movements and gait events will benefit not only diagnosis but performance assessment. With simplification of the acquisition and processing algorithms, as well as selection of appropriate wearable sensors, a platform that integrates both kinematics and sEMG for home use becomes feasible.
With this report, a simplified guide for kinematic and sEMG recording segmentation, and gait event detection was presented and fully documented. Codes and algorithms, that work with the dataset of \cite{brantley_full_2018} are made available alongside this method. To adapt to other datasets, users can easily adjust expected cadence or filter level variables, improving detection and allowing the study different walking modalities, providing an initial framework for researchers to explore and adjust to their needs and data. 

Synchronising both sEMG and kinematic information is the step forward to better characterise movement, and it is crucial that both signals start being analysed in pair. Analysing sufficient sEMG recordings in ADLs is the non-invasive key to defining the patterns of muscle activity that combine to generate movement, and should be centre stage in any mobility study. The same level of "wearable" exploration that kinematics and gait cycle has received over the past decades, needs to be pursued for the sEMG case.

With adjustable parameters, the algorithm can be adapted to further ADLs scenarios, and potentially applied to patient's smartphones, for a continuous kinematic recording. Due to its simplicity and focus to aid newcomers to the gait studies area, this algorithm can also be combined with machine learning in future iterations, to allow ad-hoc heel strike identification.

\bibliographystyle{unsrt} 
\bibliography{Biblio} 

\addtolength{\textheight}{-12cm}   

\end{document}